\documentclass[pra,aps,superscriptaddress,twocolumn,amssymb,amsmath,longbibliography]{revtex4-1}
\usepackage{graphicx}
\usepackage{theorem}
\usepackage{dcolumn}
\usepackage{bm}
\usepackage{mathrsfs}
\usepackage{setspace}

\begin{document}

\title{Torquing the Condensate: Angular Momentum Transport in Bose-Einstein Condensates by Solitonic ``Corkscrew''}

\author{Toshiaki Kanai}
\affiliation{National High Magnetic Field Laboratory, 1800 East Paul Dirac Drive, Tallahassee, Florida 32310, USA}
\affiliation{Department of Physics, Florida State University, Tallahassee, Florida 32306, USA}

\author{Wei Guo}
\email[Corresponding: ]{wguo@magnet.fsu.edu}
\affiliation{National High Magnetic Field Laboratory, 1800 East Paul Dirac Drive, Tallahassee, Florida 32310, USA}
\affiliation{Mechanical Engineering Department, Florida State University, Tallahassee, Florida 32310, USA}

\author{Makoto Tsubota}
\affiliation{Department of Physics, Osaka City University, 3-3-138 Sugimoto, Sumiyoshi-Ku, Osaka 558-8585, Japan}
\affiliation{The OCU Advanced Research Institute for Natural Science and Technology (OCARINA), Osaka City University, 3-3-138 Sugimoto, Sumiyoshi-Ku, Osaka 558-8585, Japan}
\affiliation{Nambu Yoichiro Institute of Theoretical and Experimental Physics (NITEP), Osaka City University, 3-3-138 Sugimoto, Sumiyoshi-ku, Osaka 558-8585, Japan}

\author{Dafei Jin}
\affiliation{Center for Nanoscale Materials, Argonne National Laboratory, Argonne, IL 60439, USA}

\date{\today}

\begin{abstract}
When rotating classical fluid drops merge together, angular momentum can be advected from one to another due to the viscous shear flow at the drop interface. It remains elusive what the corresponding mechanism is in inviscid quantum fluids such as Bose-Einstein condensates (BECs). Here we report our theoretical study of an initially static BEC merging with a rotating BEC in three-dimensional space along the rotational axis. We show that a soliton sheet resembling a ``corkscrew'' spontaneously emerges at the interface. Rapid angular momentum transfer at a constant rate universally proportional to the initial angular momentum density is observed. Strikingly, this transfer does not necessarily involve fluid advection or drifting of the quantized vortices. We reveal that the solitonic corkscrew can exert a torque that directly creates angular momentum in the static BEC and annihilates angular momentum in the rotating BEC. Uncovering this intriguing angular momentum transport mechanism may benefit our understanding of various coherent matter-wave systems, spanning from atomtronics on chips to dark matter BECs at cosmic scales.
\end{abstract}
\maketitle

Conservation of angular momentum can have a profound effect on the dynamics of rotating fluid systems such as cyclonic eddies in the ocean \cite{Mcewan-Nature-1976} and accretion disks surrounding stars and black holes \cite{Papaloizou-ARAA-1995, Balbus-ARAA-2003}. When rotating classical fluids merge together, the viscous shear flow at the interface can lead to the formation of vortical structures due to the Kelvin-Helmholtz instability \cite{Helmholtz-book,Kelvin-book}. Angular momentum can be advected from one fluid body to another, accompanied by the drifting of the vortical structures \cite{Landau-book}. However, for invisid quantum fluids such as low temperature Bose-Einstein condensates (BECs), little is known on what flow structures may form at the interface and how the angular momentum transfer is achieved. On the other hand, understanding the mechanism of angular momentum transport between merging rotating BECs may benefit the study of a wide range of coherent matter-wave systems. For instance, for spinning neutron stars that consist of neutron-pair superfluid \cite{Anderson-Nature-1975,Packard-PRL-1972} and for rotating galactic cold dark matter halos that are believed to form BECs \cite{Sikivie-PRL-2009}, the merging of the neutron stars \cite{Abbott-PRL-2017} and the collision of the galactic dark matter halos \cite{Zhang-EPJC-2018} may exhibit similar characteristics as merging rotating atomic BECs.

In the past, there were numerous studies on the merging dynamics of isolated atomic BECs due to its relevance to matter wave interferometry \cite{Andrews-1997-Sci,Shin-2004-PRL, Hadzibabic-2004-PRL, Shin-2005-PRA} and the celebrated Kibble-Zurek (KZ) mechanism \cite{Zurek-1996-PR, Kibble-2007-PT, Weiler-2008-Nature, Carretero-2008-PRA, Corman-2014-PRL,Lamporesi-2013-Nat}. However, most of these studies focused on merging of BECs with no initial relative motions. The merging dynamics of rotating BECs, subjecting to angular momentum conservation, is more intriguing but has received much less attention. In a recent numerical work by the authors, the merging of a rotating disk condensate with a concentric ring condensate in two-dimensional space was studied ~\cite{Kanai-2018-PRA, Kanai-2019-JLTP}. Nevertheless, since fluid advection and angular momentum transfer occurs in the same plane, it was impractical to disentangle the fluid advection effect from other possible angular-momentum transfer mechanisms.

In this paper, we report our numerical study of two cylindrical BECs merging along their rotational axis in three-dimensional (3D) space. We show that a soliton sheet resembling a ``corkscrew'' emerges at the interface of the two BECs, accompanied by rapid angular momentum transfer. Strikingly, we reveal that this transfer does not necessarily involve fluid advection or quantized vortices. Instead, a new mechanism is identified: the solitonic ``corkscrew'' can exert a torque that directly creates angular momentum in the initially static BEC and annihilates angular momentum in the rotating BEC. Uncovering this fascinating mechanism may have a far-reaching impact in various relevant topic areas.


\begin{figure}[htb]
\includegraphics[scale=0.5]{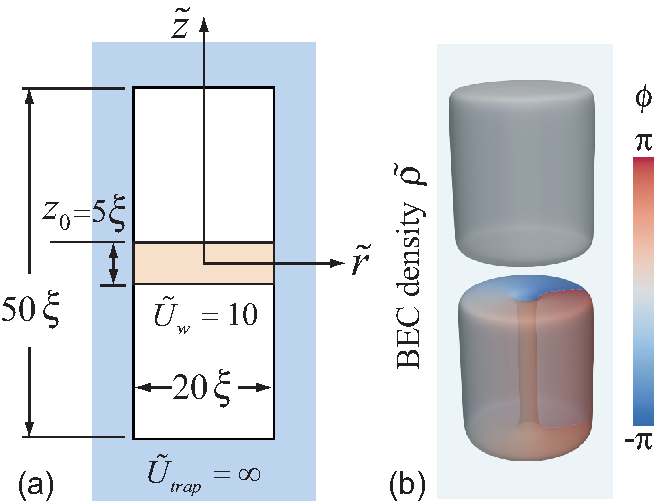}
\caption{(a) Schematic of the potential $\tilde{U}(\tilde{\textbf{r}})$ used in Eq.~\ref{Eq2}. (b) Initial density and phase profile of the BECs with a single vortex line at the center of the lower condensate. The plot shows the density iso-surface at 50\% of the bulk density.} \label{Fig1}
\end{figure}
\textbf{Numerical method:} To model the dynamical evolution of a BEC system at zero temperature, we adopt a non-linear Gross-Pitaevskii equation (GPE) \cite{Pitaevskii-2003-book}:
\begin{equation}
i\hbar \frac{\partial \psi}{\partial t}=\left[-\frac{\hbar^2}{2M}\nabla^2+U(\textbf{r},t)+g|\psi|^2\right]\psi,
\label{Eq1}
\end{equation}
where $\hbar$ is Planck's constant, $M$ is the mass of the particles that form the condensate, $\psi$=$|\psi|e^{i\phi}$ is the complex condensate wave function, $U$ is the external potential that confines the condensate, and $g$ is the coupling constant that measures the strength of the particle interactions. For convenience, we introduce dimensionless parameters $\tilde{r}$=$r/\xi$, $\tilde{t}$=$t/\tau$, and $\tilde{\psi}$=$\psi/(\sqrt{N/\xi^3})$, where $\xi$=$\hbar/\sqrt{2Mng}$ is the healing length, $\tau$=$\hbar/ng$, $N$=$\int dV|\psi|^2$ is the total particle number, and $n$=$N/V$ is the particle number density averaged over the system volume $V$. The original GPE can be written in the following dimensionless form:
\begin{equation}
i\frac{\partial \tilde{\psi}}{\partial \tilde{t}}=\left[-\tilde{\nabla}^2+\tilde{U}(\tilde{\textbf{r}},\tilde{t})+\tilde{g}|\tilde{\psi}|^2\right]\tilde{\psi}.
\label{Eq2}
\end{equation}
where the dimensionless coupling constant $\tilde{g}$ takes the form $\tilde{g}$=$N/(n\xi^3)$=$V/\xi^3$ and the dimensionless potential $\tilde{U}$=$U/ng$ now measures the ratio of the external potential $U$ to the particle interaction strength $ng$.

To study BEC merging along the rotational axis, we consider two cylindrical BECs of equal sizes that are aligned along the $\tilde{z}$-axis. This configuration is achieved by setting $\tilde{U}=\tilde{U}_{trap}+\tilde{U}_w$, where $\tilde{U}_{trap}$ represents a cylindrical hard-wall box potential that traps the condensates and $\tilde{U}_w$ denotes the potential barrier that separates the two BECs, as shown in Fig.~\ref{Fig1} (a). The hard-wall trap $\tilde{U}_{trap}$ has a diameter of $20\xi$ and a length of $50\xi$. The potential barrier, which is located at the center of the hard-wall trap, has a uniform height of $\tilde{U}_w$=10 with a thickness $\tilde{z}_0$=5. The initial state is prepared by evolving Eq.~\ref{Eq2} in imaginary time \cite{Chiofalo-2000-PRE}. Quantized vortex lines can be introduced to each condensate. An example of the initial condensate density profile with a single vortex line placed at the center of the lower condensate is shown in Fig.~\ref{Fig1} (b). At time $\tilde{t}$=0, we then suddenly remove the energy barrier $\tilde{U}_w$ and let the two condensates merge. The dynamical evolution of the condensate wavefunction can be obtained by numerically integrating Eq.~\ref{Eq2} with spatial steps $\triangle \tilde{x}$=$\triangle \tilde{y}$=$\triangle \tilde{z}$=0.2 and a time step $\triangle \tilde{t}$=4$\times10^{-5}$ using the forth-order Runge-Kutta method \cite{Press-1992-book}.

\textbf{Experimental relevance:} Our model configuration can be realized in BEC experiments. For instance, box potential has already been implemented experimentally using repulsive laser beams \cite{Navon-2016-Nature}. The size of our BEC and the height of the potential barrier are all within the parameter range of representative experiments (i.e., typical BEC size of about $10\xi$-$10^2\xi$ \cite{Kwon-2016-PRL, Scherer-2007-PRL, Jendrzejewski-2014-PRL, Corman-2014-PRL} and typical $\tilde{U}$ in the range of 1-100 \cite{Kwon-2016-PRL, Scherer-2007-PRL, Jendrzejewski-2014-PRL, Corman-2014-PRL, Killian-1998-PRL, Gaunt-2013-PRL}). It is also worthwhile noting that our BEC configuration is very similar to that used in the experiment for studying interface instability between superfluid $^{3}$He-A and $^{3}$He-B phases \cite{Blaauwgeers-2002-PRL, Volovik-2002-JETP}, although that experiment involved two immiscible superfluids and hence merging dynamics was not relevant.

\begin{figure*}[htb]
\includegraphics[scale=0.8]{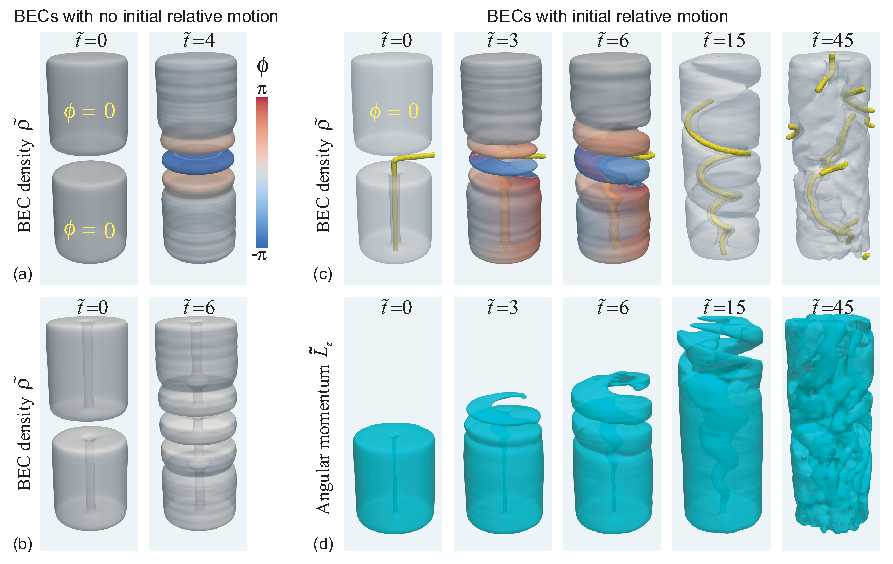}
\caption{Merging dynamics of the two condensates when they are (a) static and (b) co-rotate at $\tilde{t}$=0. (c) Condensate density evolution when only the lower condensate contains a vortex line at $\tilde{t}$=0. The color plots at $\tilde{t}$=3 and 6 show the instantaneous phase profiles. The solid yellow lines represent the locations of the vorticity singularities. (d) Evolution of the angular momentum density $\tilde{L}_z$ corresponding to (c). The plots show $\tilde{L}_z$ iso-surface at 10\% of the initial bulk value.} \label{Fig2}
\end{figure*}

\begin{figure}[htb]
\includegraphics[scale=0.57]{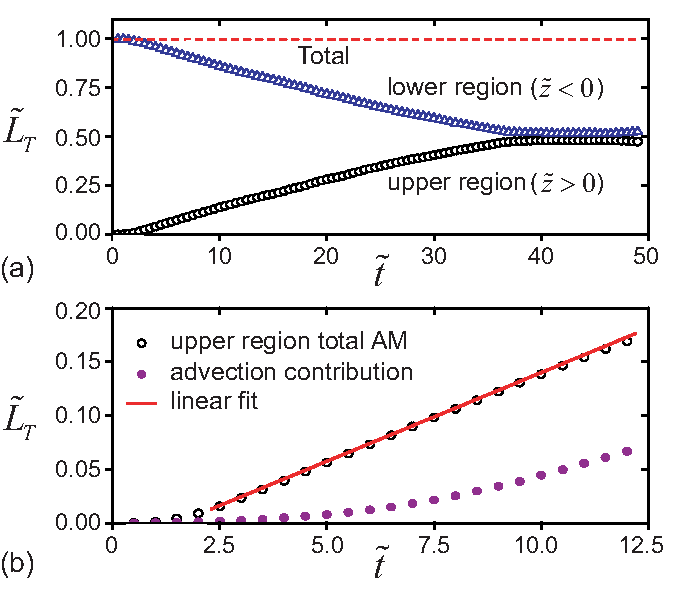}
\caption{(a) Evolution of the total angular momentum $\tilde{L}_T$ in the upper and lower condensate regions. (b) Angular momentum contribution from fluid advection in the upper condensate before vortices drift to this region.} \label{Fig3}
\end{figure}

\textbf{Simulation results:} For validation purpose, we first consider the merging of the two BECs with no initial relative motion. Fig.~\ref{Fig2} (a) and (b) show two cases where the two condensates are, respectively, static and co-rotate with a vortex line placed at the center at $\tilde{t}$=0. As the potential barrier is removed, interference fringes are formed between the two condensates. These fringes quickly evolve into disk-shaped dark solitons that propagate towards the top and bottom ends of the trap. The solitary nature of these disks is confirmed by observing the abrupt phase step $\triangle \phi$ across the density-depleted regions and the fact that they travel at the expected soliton speed $v_s$=$v_0\cdot{\texttt{cos}(\triangle\phi/2)}$ opposite to the direction of the phase step \cite{Jackson-1998-PRA}. This behavior agrees well with the observed dynamics of soliton disks created in 3D cigar-shaped condensates via phase imprint technique \cite{Shomroni-2009-NP}. In long time evolution, the soliton disks can decay into vortices due to snake instability \cite{Mamaev-1996-PRL, Theocharis-2003-PRL, Carr-book, Ma-2010-PRA}.

We now focus on BEC merging that is accompanied by angular momentum transfer. A representative case is shown in Fig.~\ref{Fig2} (c). At $\tilde{t}$=0, the lower condensate rotates around a vortex line placed at the center and carries angular momentum, while the upper condensate is static with a uniform phase $\phi$=0. Due to the phase winding in the lower condensate, the phase difference between the two condensates across the barrier gap varies around the $\tilde{z}$-axis. The time evolution of the condensate wavefunction now exhibits fascinating new features. First, a helical soliton sheet resembling a corkscrew emerges at the interface of the two condensates. This soliton sheet then extends towards both ends of the cylindrical BEC, reaches the ends at about $\tilde{t}$$\simeq$15, and bounces back, generating complex flows and density field in the merged BEC (see movies in Supplementary Materials). The solid yellow lines shown in Fig.~\ref{Fig2} (c) represent the locations of the vorticity singularities. One can see that the propagation of the helical soliton sheet in the lower condensate induces waves along the vortex line, i.e., the so-called Kelvin waves \cite{Donnelly-book, Bretin-2003-PRL, Simula-2008-PRL}. Interestingly, at relatively short evolution time (i.e., $\tilde{t}$$<$12), the vortices are nearly confined to the $\tilde{z}$$<$0 region. At long evolution times, the decay of the soliton and its interaction with the vortices and the trap wall lead to the formation of a quantum turbulence~\cite{Vinen-2002-JLTP} in the BEC that carries angular momentum, as shown in Fig.~\ref{Fig2} (c) at $\tilde{t}$=45.

To quantify angular momentum transfer, we introduce a dimensionless angular momentum density $\tilde{L}_z$ as:
\begin{equation}
\tilde{L}_z=\frac{\tau\xi}{m}\left(\psi^*\hat{L}_z\psi\right)=\frac{2}{i}\tilde{\psi}^*(\tilde{x}\frac{\partial}{\partial \tilde{y}}-\tilde{y}\frac{\partial}{\partial \tilde{x}})\tilde{\psi}.
\end{equation}
The evolution of the angular momentum density for the case presented in Fig.~\ref{Fig2} (c) is shown in (d). It seems that the angular momentum initially contained in the lower rotating condensate can rapidly ``flow'' to the upper condensate region along the helical channel formed by the soliton sheet. We can also calculate the total angular momentum $\tilde{L}_T$=$\int{\tilde{L}_zd\tilde{V}}$ integrated over the upper ($\tilde{z}>0$) and lower ($\tilde{z}<0$) condensate regions, as shown in Fig.~\ref{Fig3} (a). It appears that the rate of angular momentum transfer during merging is nearly a constant.

\begin{figure}[htb]
\includegraphics[scale=0.55]{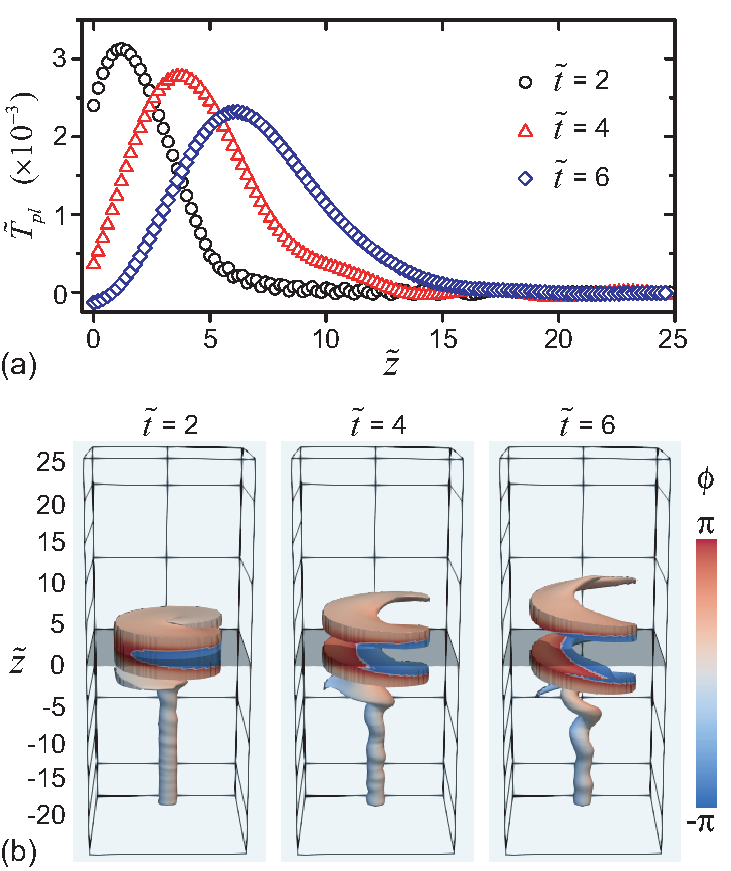}
\caption{(a) Profiles of the integrated torque exerted by the soliton sheet in the upper condensate region. (b) Profiles of the soliton sheet and the vortex line created by showing the density iso-surface at 50\% of the bulk density.} \label{Fig4}
\end{figure}

An interesting question one may raise is what the angular momentum transfer mechanism is at $\tilde{t}\lesssim12$, i.e., before the vortices drift into the upper condensate region. Fig.~\ref{Fig2} (d) may give an illusion that this transfer is controlled by fluid advection. However, this is not true. The angular momentum advected across the $\tilde{z}$=0 plane can be evaluated as $\int_0^{\tilde{t}}dt'{\int_{\tilde{z}=0}\tilde{v}_z(\tilde{r},t')\tilde{L}_z(\tilde{r},t')\cdot d^2\tilde{r}}$, where $\tilde{v}_z$=$2\partial \phi/\partial \tilde{z}$ is the velocity component along the $\tilde{z}$-axis. As shown in Fig.~\ref{Fig3} (b), this advection contribution is only a small fraction of the total angular momentum gained by the upper condensate. This result is reasonable since the flow in the lower condensate is initially perpendicular to the merging direction. The gradual increase of the advection contribution at $\tilde{t}\gtrsim7.5$ can also be understood: the Kelvin waves on the vortex line in the lower condensate deform the line into a coil shape, which then induces a vertical flow through the coil that effectively advects angular momentum to the upper region.

\begin{figure}[htb]
\includegraphics[scale=0.6]{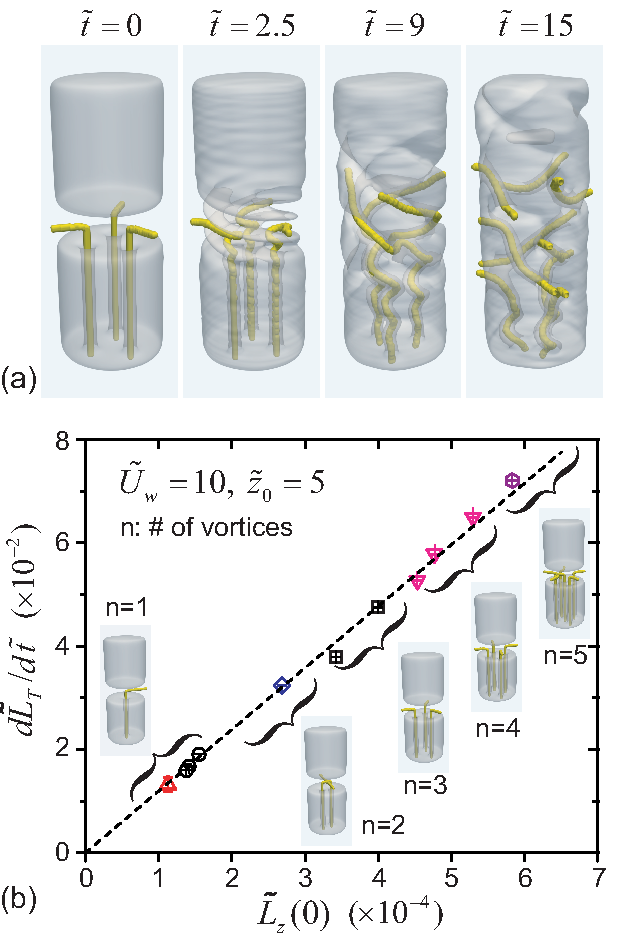}
\caption{(color online). (a) Evolution of the condensate density when the lower condensate contains three vortex lines. (b) The angular momentum transfer rate $d\tilde{L}_T/d\tilde{t}$ versus the initial angular momentum density $\tilde{L}_z(0)$ for cases with various initial vortex configurations. The barely visible error bars represent the uncertainties in the linear fit as shown in Fig.~\ref{Fig3} (b). The dashed line is a linear fit to the data.
} \label{Fig5}
\end{figure}

In order to explain the observed rapid angular momentum transfer at small evolution times, a mechanism other than advection is needed. We note that the helical soliton sheet has a free edge inside the BEC. Due to the phase step across the soliton sheet, there is a phase winding around this edge line, which induces flows in the BEC effectively like a vortex line. This feature is similar in nature to the 2D case discussed in Ref.~\cite{Kanai-2018-PRA}. As the soliton propagates, the phase profile associated with it can exert force and hence torque to both the upper and lower condensates. The torque per unit volume in the BEC with respect to the $\tilde{z}$-axis is given by:
\begin{equation}
\tilde{T}_z=(\vec{\tilde{r}}\times \vec{\tilde{f}})\cdot \hat{e}_z
\end{equation}
where the force per unit volume can be evaluated based on the change rate of the condensate momentum density: $\vec{\tilde{f}}$=$d\vec{\tilde{P}}/d\tilde{t}$=$d(|\tilde{\psi}|^2\vec{\tilde{v}})/d\tilde{t}$. To better illustrate the torque profile in the upper condensate, in Fig.~\ref{Fig4} (a) we show the total torque $\tilde{T}_{pl}$ integrated over the $\tilde{x}$-$\tilde{y}$ plane and over a step length $\triangle \tilde{z}$=0.2 along the $\tilde{z}$-axis. This torque profile moves towards the top end of the condensate just like the motion of the soliton sheet. To verify the correlation between the torque and the soliton sheet, we plot the corresponding soliton profiles in Fig.~\ref{Fig4} (b). It turns out that the spatial extend of the torque matches well with the soliton profile. The peak of the torque profile roughly coincides with the center of the soliton profile in the upper condensate region. We have also checked that the angular momentum created by the total torque in the upper condensate region matches exactly the difference between the two curves in Fig.~\ref{Fig3} (b), which therefore confirms that the torque is the missing mechanism for the angular momentum transfer. Note that the soliton sheet also exerts a torque to the lower condensate. But since the phase step of the soliton sheet reverts its direction across the $\tilde{z}$=0 plane (e.g., see Fig.~\ref{Fig2} (a)), the torque in the lower condensate has a negative sign, thereby annihilating the angular momentum in this region. The exact torque profile in the lower condensate is complicated due to the existence of the vortices, but the magnitude of the total torque matches that in the upper condensate, which warrants angular momentum conservation.

We have also examined the rate of angular momentum transfer in the early stage of BEC merging where the torque mechanism plays the key role. This rate can be determined through a linear fit to the total angular momentum data as shown in Fig.~\ref{Fig3} (b). We are curious about how this rate may depend on the initial angular momentum density difference $\tilde{L}_z(0)$ between the two condensates. To investigate this effect, we vary $\tilde{L}_z(0)$ by introducing multiple vortex lines in the lower condensate while keeping the upper condensate static. Furthermore, for a given number of vortex lines in the lower condensate, $\tilde{L}_z(0)$ can be further tuned by varying the distance between the vortices and the $\tilde{z}$-axis. Fig.~\ref{Fig5} (a) shows an example case with three vortex lines in the lower condensate at $\tilde{t}$=0. Instead of having one soliton sheet, three solitonic corkscrews emerge and twist together. A constant angular momentum transfer rate $d\tilde{L}_T/d\tilde{t}$ is again observed at short evolution times, and this indeed holds for every cases we have studied. In Fig.~\ref{Fig5} (b), we plot the obtained $d\tilde{L}_T/d\tilde{t}$ against $\tilde{L}_z(0)$ for all the cases. It is remarkable to observe that the rate $d\tilde{L}_T/d\tilde{t}$ is virtually proportional to $\tilde{L}_z(0)$ regardless of the vortex configurations. This universality may be understood qualitatively as follows. The $\tilde{L}_z(0)$ depends on the exact vortex configuration. At the meanwhile, for any vortex configuration, the solitonic corkscrews are always initiated at locations where the vortex lines are. Therefore, the spatial arrangement of the solitonic corkscrews mimics the geometric configuration of the vortex lines. The resulting total torque (which equals $d\tilde{L}_T/d\tilde{t}$ when the torque mechanism dominates) depends on this spatial arrangement in a similar way as the dependance of $\tilde{L}_z(0)$ on the vortex configuration. Therefore, the total torque (and hence $d\tilde{L}_T/d\tilde{t}$) appears to be consistently proportional to $\tilde{L}_z(0)$ at short evolution times.

In summary, our work has revealed the formation of solitonic corkscrew structures at the interface of merging rotating BECs. These corkscrews enable angular momentum transfer by exerting torques to the BECs. The rate of the angular momentum transfer appears to be universally proportional to the initial angular momentum density difference. These findings not only enrich our knowledge of BEC merging dynamics but also benefit the study of the merging dynamics of various rotating coherent matter-wave systems.

\begin{acknowledgments}
T. K. and W. G. acknowledge the support by the National Science Foundation under Grant No. DMR-1807291. The work was conducted at the National High Magnetic Field Laboratory, which is supported by National Science Foundation Cooperative Agreement No. DMR-1644779 and the state of Florida. D. J. acknowledges the support from the Center for Nanoscale Materials, an Office of Science user facility that is supported by the U.S. Department of Energy under Contract No. DE-AC02-06CH11357. M. T. was supported by JSPS KAKENHI Grant No. JP17K05548.
\end{acknowledgments}


%

\end{document}